\documentstyle[12pt]{article}
\newcommand{\nc}{\newcommand}
\nc{\neel}{N\'{e}el}
\nc{\qq}{\P}
\nc{\rng}{\rangle}
\nc{\lng}{\langle}
\nc{\rcite}{ref.~\cite}
\nc{\ba}{\begin{array}}
\nc{\ea}{\end{array}}
\nc{\Id}{I}
\nc{\reals}{{\sf I}\kern-.12em{\sf R}}
\nc{\compl}{{\sf I}\kern-.48em{\sf C}}
\nc{\lb}{\left(}
\nc{\rb}{\right)}
\nc{\qrt}{\frac{1}{4}}
\newcommand{\al}{\alpha}
\newcommand{\bt}{\beta}

\newcommand{\dl}{\delta}
\newcommand{\ep}{\epsilon}
\newcommand{\varep}{\varepsilon}

\newcommand{\sg}{\sigma}

\newcommand{\noi}{\noindent}
\newcommand{\half}{\frac{1}{2}}

\newcommand{\rf}[1]{(\ref{#1})}

\newcommand{\ra}{\rightarrow}
\newcommand{\ua}{\uparrow}
\newcommand{\da}{\downarrow}
\newcommand{\be}{\begin{equation}}
\newcommand{\ee}{\end{equation}}
\newcommand{\bea}{\begin{eqnarray}}
\newcommand{\eea}{\end{eqnarray}}
\nc{\nn}{\nonumber}
\nc{\eqrf}{eq.~\rf}
\nc{\etw}{\tilde{\eta}}

\newcommand{\pr}[1]{Phys.~Rep.~#1\ }

\begin{document}
\setlength{\unitlength}{3mm}
\newlength{\orgbaselineskip}
\setlength{\orgbaselineskip}{\baselineskip}

 \renewcommand{\thefootnote}{\fnsymbol{footnote}}

\hfill OUTP--93--18S\\
\hspace*{\fill} cond-mat/9306045\\[1cm]

\begin{center}
{\LARGE\bf
     Heisenberg models\\
     and a particular isotropic model\\[1.0cm]
}
\large
     A.J.\ van der Sijs\footnote{Supported by
     SERC grant GR/H01243.}\\[5mm]

\normalsize
     Theoretical Physics, University of Oxford\\
     1 Keble Road, Oxford OX1 3NP, United Kingdom\\[0mm]
     [e-mail: {\tt vdsijs@thphys.ox.ac.uk}]\\[1cm]


      To be published in Phys.\ Rev.\ B\\[1cm]

\end{center}

\vfill

\begin{center}
{\bf Abstract}\\
\end{center}

\noi
The Heisenberg model, a quantum mechanical analogue of the Ising model,
has a large
ground state degeneracy, due to the symmetry generated by the total spin.
This symmetry is also responsible for degeneracies in the
rest of the spectrum.
We discuss the global structure of the spectrum of Heisenberg models
with arbitrary couplings, using group theoretical methods.
The Hilbert space breaks up in blocks characterized by the
quantum numbers of the total spin, $S$ and $M$, and each block is
shown to constitute the representation space of an explicitly given
irreducible representation of the symmetric group $S_N$, consisting of
permutations of the $N$ spins in the system.

In the second part of the paper we consider, as a concrete application,
the model where each spin is coupled to all the other spins with equal
strength. Its partition function is written as a single integral,
elucidating its $N$-dependence.
This provides a useful framework for studying finite
size effects.
We give explicit results for the heat capacity, revealing
interesting behavior just around the phase transition.

\thispagestyle{empty}
\newpage
\typeout{*** Next section: Intro}
\section{Introduction}
 \setcounter{footnote}{0}
 \renewcommand{\thefootnote}{\arabic{footnote}}

The Ising model is one of the best known models of interacting spins.
In its simplest form, with nearest-neighbor interactions only,
the Hamiltonian is
\be
H_{\rm Ising} = - J \sum_{\lng ij\rng} s_i s_j   \; ,      \label{101}
\ee
where the spins take the values $+1$ (up) or $-1$ (down) and the
summation is over nearest-neighbor pairs $\langle ij \rangle$.
The ground state of the model depends on the sign of the coupling
parameter $J$.
If $J>0$ (ferromagnetic behavior) the spins tend to align themselves
with their neighbors, and the ground state is the configuration with
either all the spins up or all down.
For $J<0$ (antiferromagnet) the ground state is the checkerboard-like
\neel\ state, with all the spins up at the even
sites\footnote{A site $i=(i_1,i_2,\cdots,i_d)$ is called even (odd)
when $i_1+i_2+ \cdots+ i_d$ is even (odd).}
and down at the odd sites or the other way round.
The twofold degeneracy of the ground state in both cases reflects the
spontaneous breaking of the $Z_2$ symmetry of flipping all the spins.

The Heisenberg model is a quantum mechanical analogue of the Ising model
(for an introduction see e.g.\ ref.~\cite{affleck}, an early discussion
can be found in \rcite{vleck}).
The spectrum of this model is not as well understood, in particular the
antiferromagnetic ground state is not known.
In the Heisenberg model, however, there is a global rotation symmetry,
associated to the operator of the total spin, which is very helpful in
determining the global structure of the spectrum. This applies to the
model in its most general form, i.e.\ with arbitrary coupling between
each pair of spins.

In the first part of this paper we discuss the global
spectral structure of Heisenberg
models from a group theoretical point of view.
We start with a pedagogical introduction to the
Heisenberg model (section \rf{heisensection}),
demonstrating the role of the total-spin operator and
writing down explicitly the degenerate ferromagnetic ground states.
In section \ref{groupsection} we consider the consequences of the
global rotational symmetry for the rest of the spectrum.
It is shown that the spectrum breaks up into blocks labelled by
the quantum numbers of the total spin and
characterized by an explicitly given representation of $S_N$, the
symmetric group on $N$ elements.
Parts of this presentation are well known, but our
approach provides an attractive alternative
point of view, which exposes clearly the
$S_N$-representation structure of the spectrum.

In section \ref{examplesection} we consider the isotropic infinite
range Heisenberg model, with all the spins equally coupled.
In this model, knowledge of the global
structure of the spectrum is sufficient to write
the partition function in the form of a single integral which shows a
simple dependence on $N$. This result is particularly interesting for
studying finite size deviations from the model's thermodynamic
($N\ra\infty$) behavior. As a concrete application we present
calculations for the heat capacity.

\typeout{*** Next section: Heisenberg intro}
\section{The Heisenberg model}
\label{heisensection}

In the Heisenberg model, the spins are operators in a Hilbert space.
The Hamiltonian is
\be
H = - J \sum_{\lng ij\rng} \vec{s}_i\cdot\vec{s}_j         \label{102}
\ee
where the components $s_i^x, s_i^y, s_i^z$ of a single spin $\vec{s}_i$
constitute a set of generators of the rotation group SU(2).
The spins satisfy the commutation relations
\be
  [s_i^a,s_j^b] = i \dl_{ij} \varep^{abc} s_i^c   .         \label{103}
\ee
For a single spin, the states are labelled by the quantum numbers
$(s,m)$, determined by the eigenvalues $s(s+1)$ and $m$ of $\vec{s}^{\,2}$
and $s^z$, respectively.
We shall restrict ourselves to the spin-$\half$ case, which
bears a close resemblance to the Ising model.
We take $s^a=\sg^a/2$ where the $\sg^a$ are the
Pauli matrices
\be
\sg^x=\lb \ba{rr}0& 1\\1& 0 \ea \rb \; , \;\;
\sg^y=\lb \ba{rr}0 &-i\\i& 0 \ea \rb \; , \;\;
\sg^z=\lb \ba{rr}1& 0\\0 &-1 \ea \rb \; ,      \label{104}
\ee
in terms of which the
up state is ${\scriptsize (\ba{c} 1 \\ 0 \ea )}$, with quantum numbers
$(s=\half, m=\half)$, and the down state is ${\scriptsize
(\ba{c} 0 \\ 1 \ea )}$ with $(s=\half, m=-\half)$.

It is convenient to write the Hamiltonian as
\be
H = - J \sum_{\lng ij\rng } s_i^z s_j^z + \half (s_i^+ s_j^- + s_i^-
    s_j^+),          \label{105}
\ee
in terms of the usual ladder operators
\be
   s^{\pm} = s^x \pm i s^y      \label{106}
\ee
with the properties
\be
\ba{ccccccccc}
s^+ \lb \ba{c} 1 \\ 0 \ea \rb &=& 0 &,&\;\;\;\;\;& s^+ \lb \ba{c} 0 \\ 1
    \ea \rb
  &=& \lb \ba{c} 1 \\ 0 \ea \rb &,\\
s^- \lb \ba{c} 1 \\ 0 \ea \rb &=& \lb \ba{c} 0 \\ 1 \ea \rb &,&\;\;\;\;\;&
  s^- \lb \ba{c} 0 \\ 1 \ea \rb &=& 0 &.    \rule{0ex}{5ex}
\ea      \label{107}
\ee
The $s_i^z s_j^z$ term in \rf{105} has the same
effect as the Ising Hamiltonian \rf{101}. It tends to align the
neighboring spins $(i,j)$ (anti)parallel in the (anti)ferromagnetic
case.
The term between brackets in \rf{105}, sometimes called the
fluctuation term, is special to the quantum case.
It gives zero when acting on parallel neighboring spins, so that the
ferromagnetic Ising ground states with all spins up or all
down are also ground states of the Heisenberg ferromagnet.
Its effect on a pair of opposite spins, however,
is to flip both the spins, so that the \neel\ states
are not even eigenstates of the Heisenberg antiferromagnet.
For a two-spin system the solution is of course
to take the antisymmetric (singlet) combination
\be
\frac{1}{\sqrt{2}} \lb | \ua \da \rangle -| \da \ua \rangle \rb \; \; ,
   \label{108}
\ee
but for a system of many spins the antiferromagnetic ground
state is not known.

However, there is more to the ferromagnetic Heisenberg model as well.
Its ground state degeneracy is not just twofold but
$(N+1)$-fold, where $N$ is the number of spins in the system
(see \cite{anderson,ma}).
The reason for this large degeneracy is that the total spin
operator $\vec{S}$ commutes with the Hamiltonian, providing us with the
quantum numbers $S$ and $M$.
Since the components $S^{x,y,z}$ do not mutually commute,
one can create states with $M=-S+1, -S+2,\ldots,S$, degenerate in
energy, by repeated application of $S^+$ on a state with $M=-S$.
(Of course, the degeneracy is lifted when the global symmetry is
broken by an external magnetic field, through the addition of an
interaction term
\be
 H_B = B S^z = B \sum_i s^z_i       \label{115}
\ee
to the Hamiltonian. This causes a `fanning out' of the energy levels,
the familiar Zeeman effect.)

It is amusing to see what the non-trivial ground states actually
look like, using this ladder procedure. Starting from the
state with all spins down one obtains the states
\bea
 |\mbox{{\bf 0}}\rng &=& |\da \da \cdots \da \rangle \nn \\
 |\mbox{{\bf 1}}\rng &=& \frac{1}{\sqrt{N}} \lb |\ua \da \cdots \da \rangle
     \;\;+\;\; |\da \ua \da \cdots \da \rangle \;\;+\;\; \cdots \;\;+\;\;
     |\da \cdots \da \ua \rangle \rb    \nn \\
 |\mbox{{\bf 2}}\rng &=&  {\small \left( \ba{c} N\\ 2 \ea \right)}^{-\half}
     {\large (}\, |\ua \ua \da \da \cdots\da \rangle \;\;+\;\; |\ua\da\ua\da
     \cdots\da \rangle \;\;+\;\; \cdots \;\;+\;\; |\da \cdots\da\da\ua\ua
     \rangle \, {\large )} \nn \\ &\vdots& \nn \\
 |\mbox{{\bf N}}\rng &=& |\ua\ua \cdots\ua \rangle .   \label{109}
\eea
The state $|\mbox{{\bf K}}\rng$ is the properly normalized symmetric
sum over the ${\scriptsize \left( \ba{c} N\\ K \ea \right)}$ states
with $K$ spins up and $N-K$ spins down. Its quantum numbers are
\be
S=\half N, \;\;\;\;\;\;M=K-\half N  \;\; ,        \label{110}
\ee
in terms of which
\be
  S^{\pm} \, |\mbox{{\bf K}}\rng =
   \sqrt{(S \pm M+1)(S \mp M)} \; |\mbox{{\bf K}}\, \mbox{\boldmath
      $\pm$}\, \mbox{{\bf 1}}\rng  \; .  \label{111}
\ee

At first sight it may seem surprising that the states
$|${\bf K}$\rng$, containing antiparallel spin pairs, are ground states
of the ferromagnet.
A particularly interesting one is $|${\bf \mbox{$\half$N}}$\rng$,
a linear combination of states with as many up as down-spins
(even the \neel\ states contribute to it!).
More generally, for finite $N$ {\em each\/} of the $2^N$ basis
vectors $|\pm,\pm,\cdots,\pm\rng$ (where $+=\ua, -=\da$) has a
nonzero projection on one of the $(N+1)$ ground states of the ferromagnet.
This indicates that this set of basis vectors is not very suitable
to describe the spectrum of the Heisenberg model.
The reason is of course that the up and down states are defined with
respect to a pre-defined $z$-axis, obscuring all rotational symmetry.

To illustrate the relation with global rotational symmetry, consider
for example the expectation value of the two point function
$\langle s^z_i \, s^z_j \rangle$ in the different ground states.
For the states $|${\bf 0}$\rng$ and $|${\bf N}$\rng$ one finds
\be
\langle s^z_i \, s^z_j \rangle = \qrt \; ,   \label{112}
\ee
but the other ground states give
\be
\lng\mbox{{\bf K}}| s^z_i \, s^z_j |\mbox{{\bf K}}\rng = \qrt \lb 1 -
     4 \frac{K(N-K)}{N(N-1)} (1-\dl_{ij}) \rb  \;
     .      \label{113}
\ee
Summation over $i$ and $j$ gives $(\half
N-K)^2=M^2$, confirming the $S^z$ eigenvalue of
$|${\bf K}$\rng$.
The average of \eqrf{113} over the ground states $|${\bf K}$\rng$ is
\be
\mbox{}_{{\scriptsize \mbox{av}}}\lng s^z_i \, s^z_j
   \rng_{{\scriptsize \mbox{av}}} =
  \frac{1}{12}(1+2\dl_{ij})         .  \label{114}
\ee
For $i\neq j$, this is one third of the value $\qrt$ it attains in the
states $|${\bf 0}$\rng$ and $|${\bf N}$\rng$, as expected from rotational
invariance.

The rotational symmetry generated by the total spin
is a general feature of Heisenberg models,
it is not typical of the nearest-neighbor model with equal
interaction strengths we have been considering so far.
In fact, the entire spectrum breaks up in blocks
labelled by the quantum numbers of the total spin, each block
being characterized by a certain irreducible representation of
$S_N$, the symmetric group, or permutation group, on $N$ elements.
This is the topic of the next section.

\typeout{*** Next section: Group theory}
\section{Structure of the spectrum}
\label{groupsection}

We will start out with a general spin-$\half$ Hamiltonian for
a system of $N$ spins,
\bea
H &=&  - \sum_{\lng ij \rng} J_{ij} \, \vec{s}_i\cdot\vec{s}_j
   \label{201}\\
  &=&  - \half \sum_{\stackrel{i,j=1}{i\neq j}}^N J_{ij} \lb s_i^z s_j^z
         + \half (s_i^+ s_j^- + s_i^- s_j^+) \rb \; .
   \label{202}
\eea
The arbitrary interaction strengths $J_{ij}$ depend on $i$ and $j$
now and are not restricted to nearest-neighbor pairs. We assume
\be
J_{ij} = J_{ji} \;\;\; , \;\;\; J_{ii}=0 \;\;\;\;\mbox{for all } i,j
    \label{203}
\ee
without loss of generality.

As a consequence of the general form of this Hamiltonian,
neither the dimensionality of the problem nor the boundary conditions are
specified a priori. All the information is contained in the set of
coupling parameters $J_{ij}$.
In the nearest-neighbor model, for example, the dimension is determined
by the number of nearest neighbors of a given spin, which is the number
of non-zero couplings $J_{ij}$ for fixed $i$.
Similarly, (anti)periodic or free boundary conditions can be imposed by
choosing the $J_{ij}$ appropriately.

It is not difficult to show that $H$ is invariant under global
rotations.
The $N$-spin Hilbert space is the tensor product of $N$ single-spin
Hilbert spaces. In terms of this tensor product,
the operator for the total spin has the form
\be
\ba{rclcl}
 \vec{S} = {\displaystyle \sum_i} \vec{s}_i &=& \vec{s} \otimes
        \Id \otimes \cdots
  \otimes \Id & + & \Id \otimes \vec{s} \otimes \Id \otimes \cdots \otimes
  \Id\\ &+& \cdots & + & \Id \otimes \Id \otimes \cdots \otimes \Id \otimes
  \vec{s} \; ,
\ea   \label{204}
\ee
where the $i^{th}$ factor in each term acts on the $i^{th}$ spin.
The commutator of a single-spin component with the Hamiltonian is
\be
  [s_i^a,H] =  i \ep^{abc} \sum_j J_{ij}\, s_i^b s_j^c .    \label{205}
\ee
{}From this it is clear that
\be
[\vec{S},H] = 0        \label{206}
\ee
since each term on the right hand side of \eqrf{205} is canceled by a
similar contribution from $[s_j^a,H]$.
This means conservation of total spin $\vec{S}$.
Its components satisfy the commutation relations
\be
[S^a,S^b]=i\epsilon^{abc}\,S^c         \label{206a}
\ee
so the states in the spectrum carry the quantum numbers $S$ and $M$.

We will follow a different approach, though.
The Hamiltonian can be interpreted as an element of the group algebra
of $S_N$, the symmetric group on $N$ elements.
{}From this point of view, the Hilbert space is the representation space for
an $S_N$-representation on which $H$ acts.
This space will be shown to break up into blocks
and $S$ and $M$ will arise as natural labels on these blocks.
The states in each block constitute an irreducible representation of
$S_N$ which is given explicitly.

First, we will show that the Hamiltonian can be viewed as an element of
the group algebra of $S_N$.
The group algebra of a group $G$ is the set of formal sums $\sum_{g\in G}
x_g\cdot g$, where the $x_g$ are numbers, with the obvious addition and
multiplication rules. The group algebra plays an important role in
the representation theory of finite groups.
Now consider a term $\lng ij \rng$ in $H$,
\be
H_{\lng ij \rng} = s_i^z s_j^z + \half (s_i^+ s_j^- + s_i^- s_j^+)
         \label{207}
\ee
(ignoring the prefactor for the moment).
Its action only depends on the spins $i$ and $j$ of the state in Hilbert space
on which it acts.
Denoting the values of these spins by $|i,j\rng$, we have
\bea
H_{\lng ij \rng} |\ua\ua\rng &=& \qrt |\ua\ua\rng \; ,        \nn \\
H_{\lng ij \rng} |\da\da\rng &=& \qrt |\da\da\rng \; ,        \nn \\
H_{\lng ij \rng} |\ua\da\rng &=& \qrt {\large (} -1 |\ua\da\rng
   + 2 |\da\ua\rng {\large )}   \; , \nn \\
H_{\lng ij \rng} |\da\ua\rng &=& \qrt {\large (} -1 |\da\ua\rng
   + 2 |\ua\da\rng {\large )}   \; . \label{208}
\eea
This can be written in a general form as
\bea
H_{\lng ij \rng} |i,j\rng &=& \qrt {\large (} -1 |i,j\rng
     + 2 |j,i\rng {\large )} \nn \\
   \mbox{}&=& \qrt \lb 2 (ij) - \Id \rb |i,j\rng  \;   , \label{209}
\eea
where $(ij) \in S_N$ is the 2-cycle permutation or transposition
interchanging the spins at sites $i$ and $j$.
The contribution of the identity operator $\Id$, which is the unit
element of $S_N$, can be removed by adding a
constant to the Hamiltonian \rf{201}, and will be neglected.
The Hamiltonian is a linear combination of terms of the form \rf{209},
with coefficients proportional to $J_{ij}$,
\be
H = - \half \sum_{\lng ij \rng} J_{ij} \; (ij)   \; . \label{209a}
\ee
As such, it is an element of the group algebra of $S_N$.
Note that it is restricted to the subspace of this algebra spanned by
the conjugacy class of transpositions (2-cycles).
In the calculation of the partition function, however, one needs the
exponentiated Hamiltonian $\exp[-\beta H]$ which is not
restricted to this conjugacy class.

The $2^N$-dimensional Hilbert space of the spin model can thus be viewed
as the representation space of a representation of $S_N$.
It is clear that this representation is reducible.
Any permutation leaves the numbers of up and down-spins unchanged,
so the representation space
decomposes into a sum of invariant subspaces $W_K$ of dimension
${\scriptsize \left( \ba{c} N\\ K \ea \right)}$,
corresponding to the sectors of different $S^z$
eigenvalues $M=K-\half N$.
Each basis element of the subspace $W_K$ can be labelled by the $K$ sites
where the up-spins are located,
\bea
W_0 &=& \{()=|\mbox{{\bf 0}}\rng\}         \nn \\
W_1 &=& \{(1),(2),(3),\ldots,(N)\}         \nn \\
  &\vdots& \nn \\
W_K &=& \{\,(i_1,i_2,\ldots,i_K)\,|\,1 \leq i_1 < i_2 < \cdots < i_K \leq N\}
              \nn \\
  &\vdots& \nn \\
W_N &=& \{(1,2,3,\ldots,N)=|\mbox{{\bf N}}\rng\}    .    \label{210}
\eea
By taking the symmetric sum of all the states in $W_K$, for a fixed
value of $K$, one recovers the state $|${\bf K}$\rng$ of \eqrf{109},
which is a singlet under $S_N$.

Indeed, the representation subspaces $W_K$ are not irreducible
under $S_N$ either.
Not surprisingly, they decompose into irreducible sectors with different
total spin quantum number $S$.
This decomposition, described nicely in Wigner's book \cite{wigner},
can be summarized as follows.
Within each subspace $W_K$ (for $K\leq\half N$) there is a subset of
states which transform as $W_{K-1}$. This gives a decomposition of $W_K$
as
\be
W_K = W_{K-1} \oplus D_K        \label{210a}
\ee
where $D_K$ can be shown to be an irreducible subspace.
Subsequently, one applies the same decomposition for $W_{K-1}$ and so on
until one ends up with the full decomposition
\be
W_K = D_0 \oplus D_1 \oplus D_2 \oplus \cdots \oplus D_K \; .
\label{210b}
\ee
Finally, the decomposition of $W_K$ for $K > \half N$ is found by noting
that
\be
W_{N-K} \simeq W_K \; .  \label{210c}
\ee

Before discussing the representation spaces $D_K$, let us say something
more about this recursive decomposition of $W_K$.
Since $W_{K-1}$ can be regarded as a subspace of $W_K$ according to
\eqrf{210a}, it is interesting to see how the corresponding states
in the two spaces are mapped on each other.
For every state with $K-1$ up-spins we want to
determine the corresponding state in the space spanned by basis
vectors with $K$ up-spins.
Let $|\psi_{(K-1)}\rng = (i_1,i_2,\ldots, i_{K-1})$ be an arbitrary
state in $W_{K-1}$.
Then the corresponding state $|\psi_{(K)}\rng \in W_K$ is given by the sum
over the $N-K+1$ states $(j_1,j_2,\ldots,j_K) \in W_K$ satisfying the
condition that the set of numbers $\{j_{\al},\,\al=1,\ldots,K\}$ consists of
the numbers $\{i_{\bt},\,\bt=1,\ldots,K-1\}$ supplemented by one other number.
In other words, $|\psi_{(K)}\rng$ is the state given by the sum over
all the states for which
all the spins $i_a$ as well as one additional spin are up-spins.
This is exactly the state obtained by acting on $|\psi_{(K-1)}\rng$ with the
ladder operator $S^+$.

In terms of Young tableaux, the decomposition of $W_K$ can be visualized
as follows (see ref.~\cite{robinson}).
The reducible representation of $S_N$ on the
space $W_K$ is a permutation representation corresponding to the so-called
skew Young diagram
\be
\overbrace{
\begin{picture}(9,3)(0,1.1)
\multiput(0,2)(0,1){2}{\line(1,0){9}}
\multiput(0,2)(1,0){4}{\line(0,1){1}}
\multiput(8,2)(1,0){2}{\line(0,1){1}}
\put(3,2){\makebox(5,1){$\cdots$}}
\end{picture}
}^{N-K}
\;\;\;\;\rule[-8mm]{0mm}{8mm}
\overbrace{
\begin{picture}(7,3)(0,1.1)
\multiput(0,0)(0,1){2}{\line(1,0){7}}
\multiput(0,0)(1,0){4}{\line(0,1){1}}
\multiput(6,0)(1,0){2}{\line(0,1){1}}
\put(3,0){\makebox(3,1){$\cdots$}}
\end{picture}
}^{K}
\label{211}
\ee
which decomposes into the following
Young diagrams for irreducible representations
\bea
&&\overbrace{
\begin{picture}(9,2)(0,0.7)
\put(0,0.5){\line(1,0){9}}
\put(0,1.5){\line(1,0){9}}
\multiput(0,0.5)(1,0){4}{\line(0,1){1}}
\put(3,0.5){\makebox(3,1){$\cdots$}}
\multiput(6,0.5)(1,0){4}{\line(0,1){1}}
\end{picture}
}^{N}
\;\;\;\bigoplus\;\;\;\rule[-8mm]{0mm}{8mm}
\overbrace{
\begin{picture}(8,2)(0,0.7)
\put(0,2){\line(1,0){8}}
\put(0,1){\line(1,0){8}}
\put(0,0){\line(1,0){1}}
\multiput(0,0)(1,0){2}{\line(0,1){2}}
\multiput(2,1)(1,0){2}{\line(0,1){1}}
\put(3,1){\makebox(3,1){$\cdots$}}
\multiput(6,1)(1,0){3}{\line(0,1){1}}
\end{picture}
}^{N-1}
\;\;\;\bigoplus\;\;\;\rule[-8mm]{0mm}{8mm}
\overbrace{
\begin{picture}(7,2)(0,0.7)
\put(0,2){\line(1,0){7}}
\put(0,1){\line(1,0){7}}
\put(0,0){\line(1,0){2}}
\multiput(0,0)(1,0){3}{\line(0,1){2}}
\multiput(2,1)(1,0){2}{\line(0,1){1}}
\put(3,1){\makebox(3,1){$\cdots$}}
\multiput(6,1)(1,0){2}{\line(0,1){1}}
\end{picture}
}^{N-2}
\mbox{   } \nn \\
&&\mbox{       }\;\;\;\bigoplus\;\;\;\rule[-8mm]{0mm}{8mm}
\cdots
\;\;\;\bigoplus\;\;\;
\overbrace{
\begin{picture}(9,2)(0.0,0.7)
\put(0,0){$\underbrace{\rule[-2mm]{0mm}{2mm}
\begin{picture}(7,0.1)
\put(0,0){\line(1,0){7}}
\end{picture}
}_K
$}
\multiput(0,1)(0,1){2}{\line(1,0){9}}
\multiput(0,0)(1,0){4}{\line(0,1){2}}
\put(3,0){\makebox(3,1){$\cdots$}}
\put(3,1){\makebox(5,1){$\cdots$}}
\multiput(6,0)(1,0){2}{\line(0,1){1}}
\multiput(8,1)(1,0){2}{\line(0,1){1}}
\end{picture}
}^{N-K}     \;\;.
\label{212}
\eea
The irreducible representation\footnote{We use the same notation $D_K$
for the representation itself and for its representation space.} $D_K$ is
identified as
\be
D_K \; = \; \rule[-7ex]{0ex}{7ex}
\overbrace{
\begin{picture}(9,2)(0.0,0.7)
\put(0,0){$\underbrace{\rule[-2mm]{0mm}{2mm}
\begin{picture}(7,0.1)
\put(0,0){\line(1,0){7}}
\end{picture}
}_K
$}
\multiput(0,1)(0,1){2}{\line(1,0){9}}
\multiput(0,0)(1,0){4}{\line(0,1){2}}
\put(3,0){\makebox(3,1){$\cdots$}}
\put(3,1){\makebox(5,1){$\cdots$}}
\multiput(6,0)(1,0){2}{\line(0,1){1}}
\multiput(8,1)(1,0){2}{\line(0,1){1}}
\end{picture}
}^{N-K}           \;\; .
\label{213}
\ee

Schematically, the Hilbert space decomposes as follows
under $S_N$:
\be
\ba{lllllclllll}
D_0 &  &  &  &      &\mbox{            } &  D_0 &  &  &  &
      \rule{0ex}{6ex}         \\
D_0 & D_1 &  &  &          &             &  D_0 & D_1 &  &  &              \\
D_0 & D_1 & D_2 &  &       &             &  D_0 & D_1 & D_2 &  &           \\
&\vdots   &  &  &          &             &  &\vdots   &  &  &              \\
D_0 & D_1 & D_2 & \cdots & D_{\half (N-1)} &\;\;\;\;\;\;\;\;\mbox{and}
       \;\;\;\;\;\;\;\;&  D_0 & D_1 & D_2 & \cdots & D_{\half N} \\
D_0 & D_1 & D_2 & \cdots & D_{\half (N-1)} & &  &\vdots   &  &  &
\\
&\vdots   &  &  &          &             &  D_0 & D_1 & D_2 &  &           \\
D_0 & D_1 & D_2 &  &       &             &  D_0 & D_1 &  &  &              \\
D_0 & D_1 &  &  &          &             &  D_0 &  &  &  &                 \\
D_0 &  &  &  &             &             &      &  &  &  &
      \rule[-5ex]{0ex}{5ex}
\ea              \label{214}
\ee
for odd and even $N$, respectively.
In this diagram, the quantum number $S$ increases from right to left,
while $M$ runs vertically.
The dimension of $D_K$ is ${\scriptsize \left( \ba{c} N\\ K \ea
\right) - \left(\ba{c} N\\ K-1 \ea \right)}$, with the convention
${\scriptsize \left(\ba{c} N\\ -1 \ea \right)}=0$.
It is easily checked that the dimensions of the subspaces in this
decomposition add up to $2^N$.
In the limit
$N\ra\infty$, $D_{\half N}$ has dimension $2^N\cdot 4/N\sqrt{2\pi N}$,
while the largest $D_K$ is the one for $K=\half N - \half \sqrt N$,
whose dimension is $\sqrt{N/e}$ times as large.

This ends our discussion of the global spectral structure. To investigate
the spectrum in more detail one has to consider the representation theory of
the $S_N$ group algebra element $H$ of \eqrf{209a} in each of the
representations $D_K$.  This will depend on the specific form of the
Hamiltonian, i.e.\ on the couplings $J_{ij}$. In practical applications,
these couplings will not be arbitrary as in \eqrf{201} but related by
symmetries.
For example, the nearest-neighbor model on an $L^d$ lattice with periodic
boundary conditions and equal couplings $J_{ij}=J$ for all
nearest-neighbor pairs, has additional translational, rotational and
reflection invariance properties.
The exploitation of these extra symmetries,
regarded as a subgroup of $S_N = S_{L^d}$,
will be important for solving the model.

\typeout{*** Next section: Application}
\section{The isotropic infinite range Heisenberg model}
\label{examplesection}

As an example, we shall consider the model in which
each spin interacts with every other spin with equal
strength\footnote{For a discussion of the Ising analogue of this
model see e.g.\ \rcite{stanley}.}.
This model can be viewed as a system of N spins on the vertices of an
$\mbox{$(N-1)$}$-dimensional simplex
(the higher-dimensional generalization of a
tetraeder), interacting along its edges.
With this picture in mind, the model is effectively infinite-dimensional for
$N\ra\infty$.
The ferromagnetic version of this model was studied by Kittel and Shore
\cite{kitsho} a long time ago.
They derived expressions for the $N\ra\infty$ limit and presented numerical
results for various finite $N$. A remarkable finding was that the
phase transition develops very slowly as $N$ is increased.

Here, instead of taking the $N\ra\infty$ limit at the beginning, we
first derive an expression for the partition function in terms of
a single integral, elucidating its $N$-dependence.
Large-$N$ results can be recovered subsequently by calculating this
integral in the saddle-point approximation.
We will briefly mention the anti-ferromagnetic case too.

The partition function is
\be
Z = \frac{\mbox{Tr}\, \exp[-\bt H]}{\mbox{Tr}\, \Id}
      \label{301}
\ee
with Hamiltonian
\be
- \half J \,\sum_{i,j=1}^N \vec{s}_i \cdot \vec{s}_j  \ra H = - \half J\,
    \vec{S}^{\,2} + \half J \, \half N (\half N + 1)
    \;\;\;\;\;\;\;\;\;\;\;\;\;\;(J > 0)\; , \label{302}
\ee
where a constant has been added to normalize the ground state energy
to zero, and $\beta=1/k_BT$.

For this Hamiltonian, the energy of a
state is entirely determined by its quantum number $S$.
Calculation of the partition function thus reduces to counting the
number of states with this quantum number.
The states in the representations $D_K$ in the scheme \rf{214} have
$S=\half N-K$, and there are $2S+1$ such blocks, each consisting of
${\scriptsize \left( \ba{c} N\\ K \ea \right) - \left(
\ba{c} N\\ K-1 \ea \right)} =
{\scriptsize \left( \ba{c} N\\ \half N - S \ea \right) - \left(
\ba{c} N\\ \half N - (S+1) \ea \right)}$ states.
Hence one gets
\bea
\lefteqn{Z =
     \exp\left[ -\frac18 \bt J N(N+2)\right] \, \frac{1}{2^N} } \nn \\
     &&\times \, \sum_{S}^{\half N} (2S+1) \left\{ \left( \ba{c} N\\ \half N -
     S \ea \right) - \left(\ba{c} N\\ \half N - (S+1) \ea \right)
     \right\} \, \exp\left[ \half\bt J\, S(S+1)\right]
            \; ,    \label{303}
\eea
adopting the convention
     ${\scriptsize \left(\ba{c} N\\ -1 \ea \right) }=0$.
Here the summation runs over integers $S=0,1,\ldots,\half N$ for even $N$ and
over half integers $S=\half,\frac{3}{2},\ldots,\half N$ for odd $N$.
In the appendix it is shown that $Z$ can be rewritten in terms of an
integral,
\bea
Z &=&
2 \exp\left[-\half a(N+1)\right] \,
\frac{\partial}{\partial a} \,
\int_{-\infty}^{\infty}\!
\frac{d\eta}{\sqrt{2\pi /a(N+1)}} \,
\lb \exp\left[-\frac{1}{2}a\eta^2 \right] \, \cosh a\eta \rb^{N+1}
\label{Zfin1} \\
&=&
\exp\left[-\half a(N+1)\right] \,
\int_{-\infty}^{\infty}\!
\frac{(N+1) d\eta}{\sqrt{2\pi /a(N+1)}} \,
\eta\tanh a\eta \,
\lb \exp\left[-\half a\eta^2 \right] \, \cosh a\eta \rb^{N+1}
\label{Zfin2}
\eea
where
\be
a = \qrt \bt J \, (N+1)      \; . \label{afin}
\ee
The advantage of these expressions is that they exhibit clearly
how $Z$ depends on $N$.
It is interesting to note that the integral occurring in \eqrf{Zfin1}
is very similar to the analogous expression for the corresponding model
for classical spins, see e.g. \rcite{stanley}.

For large $N$, $\eta$ can be interpreted as the order parameter defined
by the spin fraction, in the sense that
\be
\lng \eta^2 \rng = \left\lng  \frac{S(S+1)}{\half N(\half N +1)} \right\rng
\label{etacorresp}
\ee
(up to a $1/N$ correction). Here the expectation value on the left
hand side is calculated with \eqrf{Zfin1} and the right hand side
with \eqrf{303}. This means that our $\eta$ is essentially the same as
the one defined by Kittle and Shore \cite{kitsho}, even though their
$\eta$ is a discrete quantity taking values in the interval $[0,1]$.
In fact, from \eqrf{Zfin2} we recover their result that for $a=1$
(i.e.\ $T=T_c$) the most probable value of $\eta$ is $(6/N)^{1/4}$,
and below we shall see \rf{etastar} that a similar correspondence holds
in the general case ($a\neq 1$).

We shall focus on the large-$N$ limit of the partition function and the
heat capacity $C_V$. For comparison we have displayed $C_V$ for
$N=2000$ in fig.~2, which is adapted from \rcite{kitsho}.
Guided by the prominent appearance
of the number $(N+1)$ in \eqrf{Zfin1} we perform an expansion in
$1/(N+1)$, keeping $a=\qrt\bt J(N+1)$ fixed; the corresponding
expansion in $1/N$, at fixed $a_0 = \qrt\bt JN$, is easily derived
from this.
For $(N+1)$ large, the integrand in \eqrf{Zfin1} is dominated by the
value of $\eta$ at which the exponent
\be
-\half\eta^2 +\log\cosh a\eta
\label{exp}
\ee
is maximal. This value $\etw$ is given by the familiar condition
\be
\tanh a\etw = \etw   \; ,
\label{etastar}
\ee
signalling a phase transition at $a=1$. Recall \rf{afin} that $a$ is
inversely proportional to the temperature, so $a=T_c/T$.

If $a\leq 1$ the only solution to \rf{etastar} is $\etw=0$. For $a>1$,
two additional solutions appear, and the integrand is maximal for these
non-trivial values of $\etw$, see fig.~1.
A Taylor expansion of the integrand around of \eqrf{Zfin1} around
$\eta=\etw$ now gives rise to an expansion in $1/(N+1)$. We shall give
results to lowest non-trivial order in $1/N$ only, although higher order
corrections are obtained easily.
For $a<1$ one finds
\bea
Z &=&
2 \exp\left[-\half a(N+1)\right]
\frac{\partial}{\partial a}\,
\frac{1}{\sqrt{1-a}} \lb 1
+ {\cal O}\lb\frac{1}{N}\rb \rb
\label{Zexp1} \\
&=&
\exp\left[-\half a(N+1)\right]
\frac{1}{(1-a)^{3/2}} \lb 1
+ {\cal O}\lb\frac{1}{N}\rb \rb \; ,
\label{Zexp2}
\eea
from which it follows that the lowest order correction to the value zero
of the heat capacity per spin
\be
C_V = \frac1N\bt^2\frac{\partial^2}{\partial\bt^2} \ln Z
\label{Cv}
\ee
in the high temperature phase is (for small $\tau=\frac{T-T_c}{T_c}
=\frac{1-a}{a}$)
\be
C_V^{(N)}(\tau>0) \ = \ \frac1N \lb \frac32 \frac1{\tau^2}
+{\cal O}(\tau^0) \rb + {\cal O}\lb \frac 1{N^2} \rb \; .
\label{Cv1}
\ee

For $a>1$, expansion of the integrand around $\pm \etw$ gives
\be
Z =
  4 \exp\left[-\half a(N+1)\right]
  \frac{\partial}{\partial a} \,
  \frac{\lb\exp\left[-\frac{1}{2}a\etw^2 \right] \, \cosh a\etw
  \rb^{N+1}}{\sqrt{1-a(1-\etw^2)}}
  \lb 1+ {\cal O}\lb\frac1N \rb \rb \, ,
\label{Zexp4}
\ee
which can be evaluated using
\be
\frac{d\etw}{da} = \frac{\etw(1-\etw^2)}{1-a(1-\etw^2)} \, .
\label{etaprime}
\ee
For small $|\tau|=\frac{T_c-T}{T_c}=\frac{a-1}{a}$ this gives for
the heat capacity in the low temperature phase
\be
C_V^{(N)} (\tau<0) \ = \ \frac32 - \frac{12}{5}|\tau| + {\cal O}(|\tau|^2)
 -\frac1N \lb \frac{1}{|\tau|^2} + {\cal O}(|\tau|^0) \rb + {\cal O}
\lb \frac {1}{N^2} \rb \; .
\label{Cv2}
\ee
This behavior of the heat capacity (\ref{Cv1},\ref{Cv2}) seems compatible
with the finite size data of \rcite{kitsho}, see also fig.~2.

These results are valid for $a$ not too close to 1, in fact we must have
$|a-1| \approx |\tau| \gg 1/\sqrt N$,
otherwise the derivations of eqs.~\rf{Zexp2}
and \rf{Zexp4} and the subsequent results for the heat capacity lose validity.
In the thermodynamic limit $N\ra\infty$, this means that only the phase
transition point $a=1$ itself is not described by these formulas, but
for finite $N$ there is a finite interval of $a$-values around $a=1$ in which
the theory behaves qualitatively different. The width of this interval
is ${\cal O}\lb\frac{1}{\sqrt N}\rb$ and can be regarded as a measure
of the ``width of the phase transition'' for finite $N$ (for convenience
we keep using the term `phase transition' although strictly speaking
there is no phase transition in a finite system).

This region very close to the phase transition appears to show
interesting behavior. Consider expression \rf{Zfin2} for the partition
function with $\tau=\frac{1-a}{a} = \frac{\dl}{\sqrt{N+1}}$,
where $0\le |\dl| \ll 1$.
The leading terms in the exponent of the integrand are
\be
(N+1) \lb - \half \frac{\dl}{\sqrt{N+1}} \eta^2 -\frac1{12}a^4\eta^4 \rb
\label{expdom}
\ee
which both appear to be of leading order in $N$, after making the
substitution $a\eta\ra y/(N+1)^{1/4}$. Since $|\dl| \ll 1$, however, we can
consider the $\dl$-dependent exponential as a perturbation. We find
\bea
Z &=&
\exp\left[-\half a(N+1)\right] \,
\frac{(N+1)^{3/4}}{a\sqrt{2\pi a}}\,
\int_{-\infty}^{\infty}\! dy \, y^2
\exp\left[ -\half \dl y^2 -\frac1{12} y^4 \right] \,
\lb 1 + {\cal O}\lb\frac1{\sqrt{N+1}}\rb\rb
\label{Zpt1} \\
&=&
\exp\,\left[-\half a(N+1)\right] \,
\frac{(N+1)^{3/4}}{a\sqrt{2\pi a}}
\nn \\
&&\mbox{} \times \,
\lb 6 (12)^{-1/4} \Gamma(\frac34)
-\half\dl\,6(12)^{1/4}\Gamma(\frac54) +\frac18 \dl^2\,6(12)^{3/4}
\Gamma(\frac74) -\frac1{48}\dl^3\, 6(12)^{5/4} \Gamma(\frac94) + {\cal
O}(\dl^4) \rb
\nn \\
&&\mbox{} \times \,
\lb 1 + {\cal O}\lb\frac1{\sqrt{N}}\rb\rb\, .
\label{Zpt2}
\eea
This leads to
\bea
C_V^{(N)} (|\tau|\ll\frac1{\sqrt N})
&=& 2\lb\frac98 -\frac32 \frac{\Gamma^2(\frac54)}
{\Gamma^2(\frac34)}\rb
- 6\sqrt 3 \frac{\Gamma(\frac54)}{\Gamma(\frac34)}
\lb\frac{\Gamma^2(\frac54)}{\Gamma^2(\frac34)}-\half\rb \dl + {\cal
O}(\dl^2) + {\cal O}\lb\frac{1}{\sqrt{N}}\rb \ \ \mbox{}
\label{Cv3} \\
&=& 0.61 - 0.36 \sqrt{N} \,\tau \ \mbox{ + higher order corrections} \; .
\label{Cv4}
\eea
Thus we have explicitly found the value of the heat capacity
and its slope {\em at\/} the phase transition in a finite system.
The fact that the slope is proportional to $\sqrt N$ confirms the
earlier remark that the width of the phase transition is of order
$\frac1{\sqrt N}$.  The behavior \rf{Cv4} appears to be in qualitative and
quantitative accordance with the numerical results for $C_V$ of
Kittle and Shore \cite{kitsho}, see fig.~2 (recall $-\tau\approx a-1$).

For completeness we mention the antiferromagnetic case (see \eqrf{310} in
the appendix) as well, although it is not very interesting.
It might seem appropriate to redefine the normalization of the
Hamiltonian \rf{302} such that the antiferromagnetic ground state energy
becomes zero, but we stick to the original normalization for continuity
reasons.
The integrand is maximal at $\etw=0$ and the partition function
becomes (here $a=-\qrt\bt J(N+1)>0$)
\be
Z= \exp\left[\half a(N+1)\right]
\frac{1}{(1+a)^{3/2}} \lb 1
+ {\cal O}\lb\frac{1}{N}\rb \rb \; ,
\label{ZAF}
\ee
which is just the ferromagnetic result \rf{Zexp2} with $a\ra -a$.
The partition function has continuous behavior around $a=0$ so we can
regard the antiferromagnetic case as an extension of the high
temperature phase of the ferromagnet.

\section{Conclusion}
\typeout{*** New section: Conclusion}

In the first part of this paper we have discussed Heisenberg
models with arbitrary couplings.
The global structure of the spectrum of such models
is determined by the well known
global rotational symmetry provided by the total spin, which causes
the spectrum to break up into blocks of states labelled by the
quantum numbers $S$ and $M$ of the total spin.
The group theoretical approach pursued here shows that the
states within one such block transform according to explicitly
given irreducible representations of the symmetric group $S_N$.

In an attempt to diagonalize the Hamiltonian in the representations $D_K$,
one might try to proceed by applying representation theory for the
$S_N$ group algebra, in terms of the Schur functions for example.
In practice, one is usually interested in Heisenberg models satisfying
extra relations between the coupling parameters $J_{ij}$, such as the
nearest-neighbor model with equal couplings between each pair of
neighboring spins.
Consideration of the additional symmetries following from these
relations, regarded as subgroups of $S_N$, is probably essential.
Another possible extension of this work would be a generalization
to spins in higher-dimensional
representations of SU(2). There, one expects the appearance of
$S_N$ representations other than the ones discussed here.
Furthermore, it would be interesting to give a description of spin waves
in the present framework.

In the second part of this paper we studied the infinite range Heisenberg
model with equal couplings between all the $N$ spins.
We have presented a compact formulation of the partition function
which clearly exhibits the dependence on $N$ and allows for a
straightforward study of deviations from the $N\ra\infty$ behavior.
As an example, we calculated finite size corrections to
the heat capacity. These calculations reveal interesting behavior
close to the phase transition and are in quantitative agreement
with numerical data for finite systems.

\appendix

\section{Rewriting the partition function}
\typeout{*** New section: Appendix}

In this appendix we present the derivation of the integral
representation (\ref{Zfin1},\ref{Zfin2}) of the partition function
\eqrf{303},
\bea
\lefteqn{Z =
     \exp\left[ -\frac18 \bt J N(N+2)\right] \, \frac{1}{2^N} } \nn \\
     &&\times \sum_{S}^{\half N} (2S+1) \left\{ \left( \ba{c} N\\ \half N -
     S \ea \right) - \left(\ba{c} N\\ \half N - (S+1) \ea \right)
     \right\} \, \exp\left[ \half\bt J\, S(S+1)\right] \; .
  \label{A1}
\eea
We will concentrate on the ferromagnetic case, $J>0$. After that we will
summarize the results for the antiferromagnet and briefly discuss the
inclusion of an external field.

The negative terms in \rf{A1} can be rewritten using the equality
\bea
\lefteqn{-(2S+1)\lb \ba{c} N\\ \half N - (S+1) \ea \right)\,
    \exp\left[ \half\bt J \, S(S+1)\right]}
    \ \ \ \ \ \ \ \ \ \ \ \ \ \ \nn \\
    && = (2S'+1)\lb \ba{c} N\\ \half N - S' \ea
    \right)\, \exp\left[ \half\bt J \, S'(S'+1)\right]         \label{304}
\eea
where $S'=-(S+1)$ and we have used that
  ${\scriptsize \left(\ba{c} N\\ \half N + S' \ea \right) =
                \left(\ba{c} N\\ \half N - S' \ea \right) }$.
Eq.~\rf{303} now reduces to the rather elegant expression
\bea
\lefteqn{Z =
     \exp\left[ -\frac18 \bt J (N+1)^2 \right] }
     \ \ \ \ \nn \\
   &&\ \times\,\frac{1}{2^N}\,\sum_{S=-\half N}^{\half N} (2S+1) \left( \ba{c}
     N\\ \half N - S \ea \right) \, \exp\left[\half\,(\qrt\bt J) \,
     (2S+1)^2\right] \, .
\label{305a}
\eea
We rewrite the last exponential by applying
\be
\exp\left[\half q p^2\right] = \frac{1}{\sqrt{2\pi q}} \,
\int_{-\infty}^{\infty}\!dt \, \exp\,\left[-\frac{1}{2q} t^2 + pt\right]
\label{306}
\ee
for $p=2S+1,\; q=\qrt\bt J$, and obtain
\bea
\lefteqn{Z=
\exp\,\left[-\half a(N+1)\right] \,\int_{-\infty}^{\infty}\!
\frac{dt}{\sqrt{2\pi a/(N+1)}}} \nn \\
&&\ \ \ \ \ \times
\exp\,\left[-\frac{1}{2a}(N+1)t^2 \right] \,
\frac{\partial}{\partial t} \frac{1}{2^N}\,
\sum_{S=-\half N}^{\half N} \left( \ba{c} N\\ \half N -
    S \ea \right) \, \exp\,[(2S+1)\, t]  \; ,
\label{307}
\eea
where we have introduced
\be
a = \qrt \bt J \, (N+1)      \; . \label{307a}
\ee
Now the summation over $S$ can be carried out, leading to
\be
Z=
\exp\,\left[-\half a(N+1)\right] \,\int_{-\infty}^{\infty}\!
\frac{dt}{\sqrt{2\pi a/(N+1)}} \,
\exp\,\left[-\frac{1}{2a}(N+1)t^2 \right] \; \frac{\partial}{\partial t}
\left( \exp t \, \cosh^N t \right) \; .  \label{308}
\ee
Upon using the $t\ra -t$ symmetry of the integrand, introducing
an additional derivative, and performing partial integration, we get
\be
Z =
\exp\left[-\half a(N+1)\right] \,\int_{-\infty}^{\infty}\!
\frac{dt}{\sqrt{2\pi a/(N+1)}} \,
\cosh^{N+1} t \;
\frac{1}{N+1} \frac{\partial^2}{\partial t^2}
\exp\left[-\frac{1}{2a}(N+1)t^2 \right] .\;\;\mbox{}  \label{308b}
\ee
With the equality
\be
\frac{\partial^2}{\partial t^2} \,
\frac{\exp\,\left[-\frac{1}{2a}(N+1)t^2 \right] }{\sqrt{2\pi a/(N+1)}}
= 2(N+1)\frac{\partial}{\partial a} \,
\frac{\exp\,\left[-\frac{1}{2a}(N+1)t^2 \right] }{\sqrt{2\pi a/(N+1)}}
\label{308c}
\ee
this turns into
\bea
Z &=&
2 \exp\,\left[-\half a(N+1)\right] \,
\frac{\partial}{\partial a} \,
\int_{-\infty}^{\infty}\!
\frac{dt}{\sqrt{2\pi a/(N+1)}} \,
\lb \exp\,\left[-\frac{1}{2a}t^2 \right] \, \cosh t \rb^{N+1}
\label{308d1} \\
&=&
\exp\,\left[-\half a(N+1)\right] \,
\int_{-\infty}^{\infty}\!
\frac{(N+1) d\eta}{\sqrt{2\pi /a(N+1)}} \,
\eta\tanh a\eta \,
\lb \exp\,\left[-\half a\eta^2 \right] \, \cosh a\eta \rb^{N+1} .
\label{308d3}
\eea
(Eq.~\rf{308d3} is obtained by computing $\partial/\partial a$ after
substituting $t\ra t'\sqrt{a}$ in \eqrf{308d1}.)

In the antiferromagnetic case, $J<0$, the derivation is the same up to
\eqrf{305a}. Instead of \eqrf{306} we then apply
\be
\exp\left[-\half q p^2\right] = \frac{1}{\sqrt{2\pi q}} \,
\int_{-\infty}^{\infty}\!dt \, \exp\,\left[-\frac{1}{2q} t^2 + ipt\right]
\label{309}
\ee
and the remainder of the derivation proceeds analogously to the
ferromagnetic case, leading to
\be
Z =
-2 \exp\,\left[\half a(N+1)\right] \,
\frac{\partial}{\partial a} \,
\int_{-\infty}^{\infty}\!
\frac{dt}{\sqrt{2\pi a/(N+1)}} \,
\lb \exp\,\left[-\frac{1}{2a}t^2 \right] \, \cos t \rb^{N+1}
\label{310}
\ee
(where now $a=-\qrt\beta J(N+1)>0$). This expression is related to
\rf{308d1} by $a\ra -a,\;t\ra it$.

For completeness we also mention the inclusion of an external field.
If we add a coupling term $-BS_3$
to the Hamiltonian, the factor $(2S+1)$ in \eqrf{A1} is replaced by
\be
\sum_{M=-S}^S \exp[-\beta BM] = \frac{\sinh \half\beta B(2S+1)}{\sinh
\half\beta B} \; .
\label{A2}
\ee
Proceeding along the lines of the derivation given above one obtains
(with $t_0 = \half\beta B$)
\bea
\lefteqn{Z = \frac{\exp\left[-\half a(N+1)\right]}{\sinh t_0}
\frac{1}{N+1}} \nn \\
&& \times\, \frac{\partial}{\partial t_0}
\, \int_{-\infty}^{\infty} \frac{dt}{\sqrt{2\pi a/(N+1)}}
\lb \exp\left[-\frac{1}{2a}t^2\right] \cosh(t+t_0)\rb^{N+1}
\label{A3}
\eea
in the ferromagnetic case, with a similar result for the antiferromagnet.

\typeout{*** References...}

\vfill

\noi
{\Large\bf Figure captions}\\[2cm]
\noi
Figure 1: The integrand in \eqrf{Zfin2} for $a<1$ (solid line)
and $a>1$ (dashed line).\\[1cm]
Figure 2: Heat capacity per spin for the isotropic model ($N=2000$) as a
function of $a=\qrt\bt J(N+1)$ (adapted from \rcite{kitsho}).
The dotted line is $\frac32 - \frac{12}{5}(a-1)$, see \eqrf{Cv2}, the
dash-dotted line is \eqrf{Cv4}: $\ 0.61 + 0.36\sqrt{2000}\,(a-1)$.
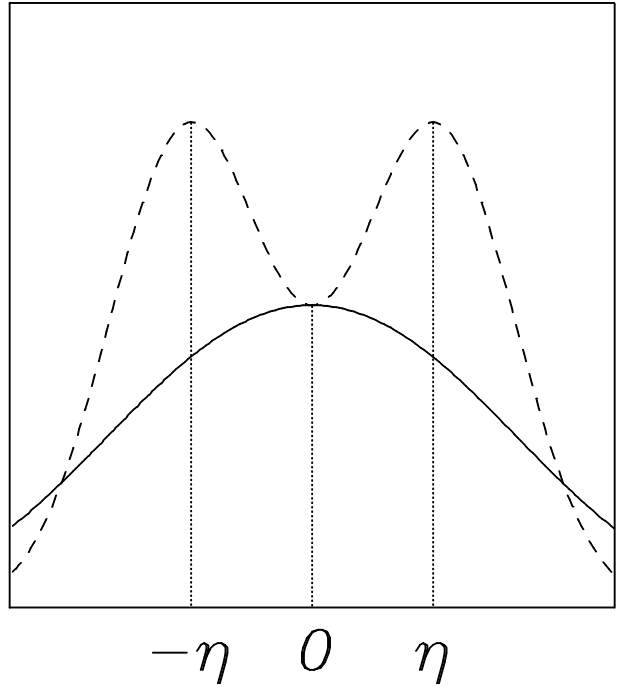
\end{document}